\documentclass{iaus}
\usepackage{graphicx}

\def\aa{\textit{A\&A}\ }
\def\annrev{\textit{ARA\&A}\ }
\def\apj{\textit{ApJ}\ }
\def\nat{\textit{Nature}\ }
\def\sci{\textit{Science}\ }
\def\gcn{\textit{GCN}\ }
\def\mnras{\textit{MNRAS}\ }
\def\lsim{\mathrel{\rlap{\lower 4pt \hbox{\hskip 1pt $\sim$}}\raise 1pt
\hbox {$<$}}} 
\def\gsim{\mathrel{\rlap{\lower 4pt \hbox{\hskip 1pt $\sim$}}\raise 1pt
\hbox {$>$}}}
\def\ion#1#2{{\rm #1}~{\sc #2}}

\newcommand{\Msun}{M_\odot}
\newcommand{\Mms}{M_{\rm MS}}
\newcommand{\Mej}{M_{\rm ej}}

\newcommand{\Cofs}{$^{56}$Co}
\newcommand{\Nifs}{$^{56}$Ni}
\newcommand{\Mni}{M{\rm (^{56}Ni)}}

\newcommand{\KE}{E_{\rm K}}

\title[First Stars -- Type Ib Supernovae Connection] 
{First Stars -- Type Ib Supernovae Connection}

\author[Ken'ichi Nomoto {\textit et al.}]   
{Ken'ichi~Nomoto$^{1,2}$, Masaomi~Tanaka$^{2,1}$, Yasuomi~Kamiya$^{2,1}$,
 Nozomu~Tominaga$^3$, and Keiichi Maeda$^1$}

\affiliation{$^1$Institute for the Physics and Mathematics of the
Universe, University of Tokyo, Kashiwa, Chiba 277-8568, Japan \\
email: {\tt nomoto@astron.s.u-tokyo.ac.jp} \\[\affilskip]
$^2$Department of Astronomy, University of Tokyo, Bunkyo-ku, Tokyo
113-0033, Japan \\
$^3$National Astronomical Observatory, Mitaka, Tokyo 113-0033, Japan}

\pubyear{2008}
\volume{255}  
\pagerange{182--188}
\jname{Low-Metallicity Star Formation: From the First Stars to Dwarf Galaxies}
\editors{L.K. Hunt, S. Madden \& R. Schneider, eds.}

\begin{document}

\maketitle

\begin{abstract}
The very peculiar abundance patterns observed in extremely metal-poor
(EMP) stars can not be explained by ordinary supernova
nucleosynthesis but can be well-reproduced by nucleosynthesis in
hyper-energetic and hyper-aspherical explosions, i.e., Hypernovae
(HNe).  Previously, such HNe have been observed only as Type Ic
supernovae.  Here, we examine the properties of recent Type Ib
supernovae (SNe Ib).  In particular, SN Ib 2008D associated with the
luminous X-ray transient 080109 is found to be a more energetic
explosion than normal core-collapse supernovae.  We estimate that the
progenitor's main sequence mass is $\Mms =$ 20--25~$\Msun$ and a kinetic
energy of explosion is $\KE \sim 6 \times 10^{51}$ erg.  These
properties are intermediate between those of normal SNe and hypernovae
associated with gamma-ray bursts.  Such energetic SNe Ib
can make important contribution to the chemical enrichment in the
early Universe.

\end{abstract}

\firstsection 
\section{Metal Poor Stars - Hypernovae - GRB Connections}

The abundance patterns of the extremely metal-poor (EMP) stars are
good indicators of supernova (SN) nucleosynthesis, because the Galaxy
was effectively unmixed at [Fe/H] $< -3$.  Thus they could provide
useful constraints on the nature of First Supernovae and thus First
Stars.

The EMP stars are classified into three groups according to [C/Fe]
(e.g., \cite{hill2005}; \cite{beers2005}):

(1) [C/Fe] $\sim 0$, normal EMP stars ($-4<$ [Fe/H] $<-3$); 

(2) [C/Fe] $\gsim+1$, Carbon-enhanced EMP (CEMP) stars ($-4<$ [Fe/H]
    $<-3$); 

(3) [C/Fe] $\sim +4$, hyper metal-poor (HMP) stars ([Fe/H] $<-5$,
    e.g., HE~0107--5240, \cite{chr02,bes05}; HE~1327--2326,
    \cite{fre05}).

In addition, Table 1 summarizes other abundance features of various EMP
stars.  Many of these EMP stars have high [Co/Fe].

We have shown that such peculiar abundance patterns can not be
explained by conventional normal supernova nucleosynthesis but can be
reproduced by nucleosynthesis in hyper-energetic and hyper-aspherical
explosions, i.e., Hypernovae (HNe) (e.g., \cite{mae02}; \cite{mae03};
\cite{tom08}; \cite{tom09}).

\begin{figure}[t]
\begin{center}
\includegraphics[width=10cm]{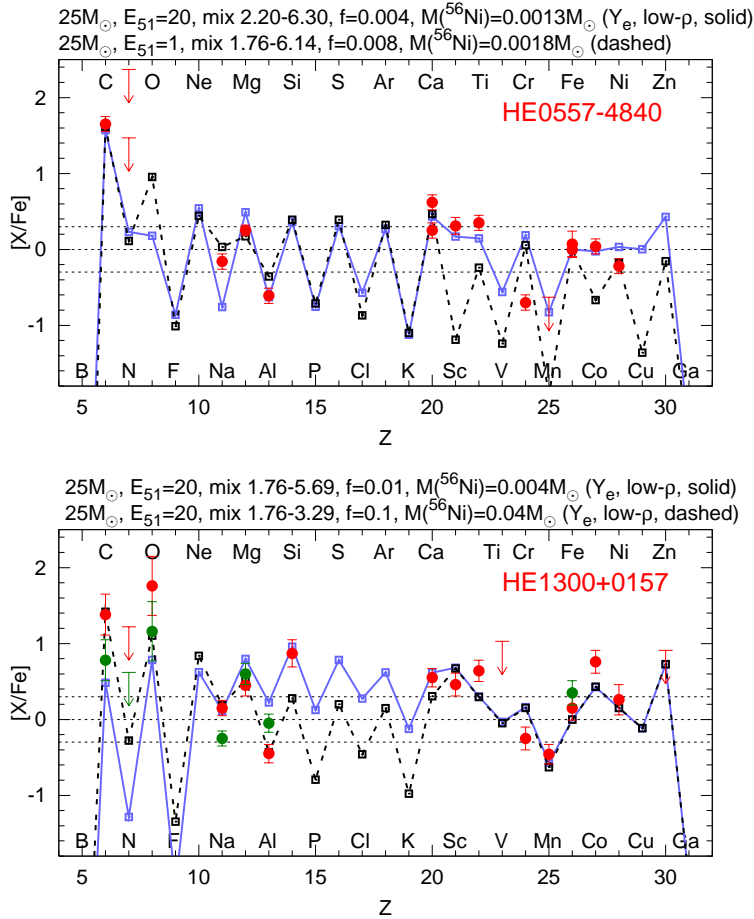}
\caption{Comparisons of the abundance patterns between the
 mixing-fallback models and the UMP star HE0557--4840 (upper:
 \cite{nor07}), and the CEP star HE1300+0157 (lower: \cite{fre07}).
}
\label{fig:UMP}
\end{center}
\end{figure}

The abundance pattern of the Ultra Metal-Poor (UMP) star
(HE~0557--4840: \cite{nor07}) is shown in Figure~\ref{fig:UMP} and
compared with the HN ($E_{51}=E_K/10^{51}$ erg $=20$, where $\KE$ is
the kinetic energy of explosion) and SN ($E_{51}=1$) models of
the 25 $\Msun$ stars.  The Co/Fe ratio ([Co/Fe] $\sim0$) requires a high
energy explosion and the high [Sc/Ti] and [Ti/Fe] ratios require a
high-entropy explosion.  The HN model is in a
good agreement with the abundance pattern of HE~0557--4840. The model
indicates $\Mni\sim10^{-3}$ $\Msun$ being similar to faint SN models for
CEMP stars.  

The abundance pattern of the CEMP-no star (i.e., CEMP with no neutron
capture elements) HE~1300+0157 (\cite{fre07}) is shown in
Figure~\ref{fig:UMP} (lower) and marginally reproduced by the
hypernova model with $\Mms=25$ $\Msun$ and $E_{51}=20$.  The large
[Co/Fe] particularly requires the high explosion energy.

Previously, Hypernova-like explosions with $E_{51} > 10$ have been
found only in Type Ic supernovae (SNe Ic), which are core-collapse
supernova and characterized by the lack of hydrogen and helium.

Recently, several interesting Type Ib supernovae (SNe Ib) have been
observed to show quite peculiar features.  SNe Ib are another type of
envelope-stripped core collapse SN but characterized by the presence
of prominent He lines.  Thus it is interesting to examine the
explosion energy and other properties of SNe Ib in comparison with
Hypernovae and normal SNe.

Here we present our analysis of peculiar SNe Ib 2008D and 2006jc
to obtain their features (\cite{tan09}; see also, \cite{maz08}).

\begin{table}
  \caption{Metal-poor stars.}
  \label{tab:stars}
  \begin{tabular}{ccccccccc}
   \hline
    Name & [Fe/H] & Features & Reference \\
   \hline
   HE~0107--5240 & $-5.3$ & C-rich, Co-rich?, [Mg/Fe] $\sim0$ &
    \cite{chr02} \\
   HE~1327--2326 & $-5.5$ & C, O, Mg-rich & \cite{fre05,aok06} \\
   HE~0557--4840 & $-4.8$ & C, Ca, Sc, Ti-rich, [Co/Fe] $\sim0$
 & \cite{nor07} \\
   HE~1300+0157 & $-3.9$ & C, Si, Ca, Sc, Ti, Co-rich
 & \cite{fre07} \\
   HE~1424--0241 & $-4.0$ & Co, Mn-rich, Si, Ca, Cu-poor
 & \cite{coh07} \\
   CS~22949--37 & $-4.0$ & C, N, O, Mg, Co, Zn-rich
 & \cite{dep02} \\
   CS~29498--43 & $-3.5$ & C, N, O, Mg-rich, [Co/Fe] $\sim0$
 & \cite{aok04} \\
   BS~16934--002 & $-2.8$ & O, Mg-rich, C-poor
 & \cite{aok07} \\
   \hline
  \end{tabular}
\end{table}

\begin{figure}[t]
\begin{center}
\includegraphics[width=8.5cm]{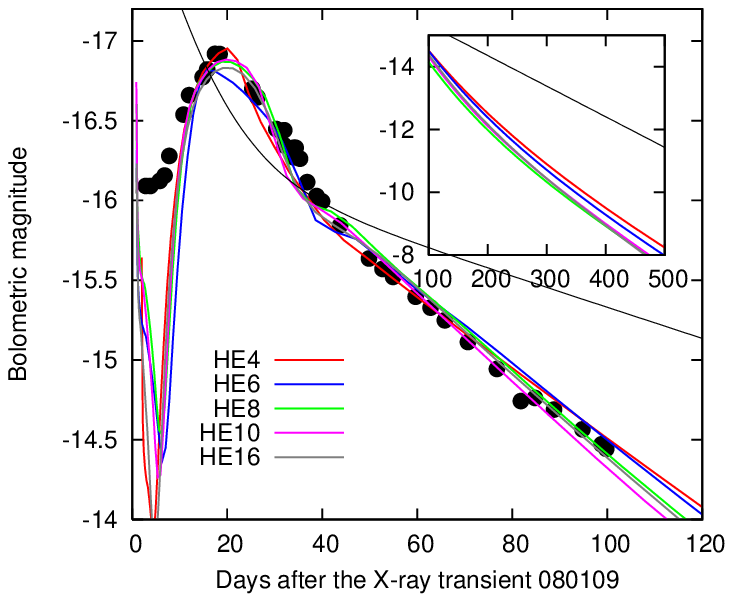}
\caption{
The pseudo-bolometric ($UBVRIJHK$) light curve (LC) of SN 2008D
compared with the results of LC calculations with the models HE4
(red), HE6(blue), HE8 (green), HE10 (magenta) and HE16 (gray)
(\cite{tan09}).  The pseudo-bolometric LC is shown in filled (left)
and open (right) circles.  The thin black line shows the decay energy
from \Nifs\ and \Cofs\ [$\Mni$ $=$ 0.07 $\Msun$].  At late epochs, it
is roughly equal to the optical luminosity under the assumption that
$\gamma$-rays are fully trapped.  The bolometric magnitude at $t \sim
4$ days after the X-ray transient is brighter by $\sim$0.25 mag than
that shown by other papers.
}
\label{fig:sn08dLC}
\end{center}
\end{figure}

\begin{figure}[t]
\begin{center}
\includegraphics[width=8.5cm]{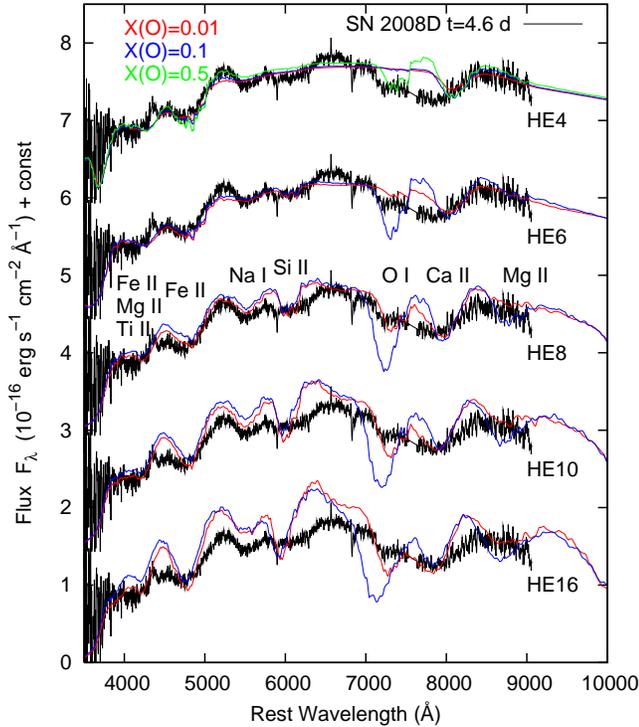}
\caption{
The spectrum of SN 2008D at $t=4.6$ days from the X-ray transient
(black line, \cite{maz08}) compared with synthetic spectra (color
lines, \cite{tan09}).  The spectra are shifted by 6.0, 4.5, 3.0, 1.5,
0.0 from top to bottom.  The model spectra are reddened with
$E(B-V)=0.65$ mag.  From top to bottom, the synthetic spectra
calculated with HE4, H46, HE8, HE10 and HE16 are shown.  The red, blue
and green lines show the synthetic spectra with oxygen mass fraction
$X$(O) = 0.01, 0.1, and 0.5, respectively.  Since the synthetic
spectra with $X$(O) = 0.1 for more massive models than HE4 already
show too strong \ion{O}{i} line, the spectra with $X$(O) = 0.5 are not
shown for these models.
}
\label{fig:sn08dSP}
\end{center}
\end{figure}

\section{Energetic Type Ib Supernova SN 2008D}

SN 2008D was discovered as a luminous X-ray transient in NGC 2770.
The X-ray emission of the transient reached a peak $\sim$65 seconds,
lasting $\sim$600 seconds, after the observation started.
SN 2008D showed a broad-line optical spectrum at early epochs ($t
\lsim 10$ days, hereafter $t$ denotes time after the transient, 2008
Jan 9.56 UT, \cite{mod08}).  Later, the spectrum changed to
that of normal SN Ib (\cite{sod08}; \cite{mal08}; \cite{mod08};
\cite{maz08}).

We have done detailed theoretical study of emissions from SN 2008D.
The bolometric LC and optical spectra are modeled based on realistic
progenitor models and the explosion models obtained from
hydrodynamic/nucleosynthetic calculations (\cite{tan09}).

The pseudo-bolometric ($UBVRIJHK$) light curve (LC) is compared with
the He star models HE4, HE6, HE8, HE10, and HE16, whose masses are
$M_{\alpha} =$ 4, 6, 8, 10, and 16~$\Msun$, respectively (Fig. 2).
These He stars correspond to the main-sequence masses
of $\Mms \sim$ 15, 20, 25, 30, and 40 $\Msun$ stars
(\cite{nom88}).

Since the timescale around the peak depends on both the ejected mass
$\Mej$ and $\KE$ as $\propto \kappa^{1/2} \Mej^{3/4} \KE^{-1/4}$,
where $\kappa$ is the optical opacity (\cite{arn82}), a specific
kinetic energy is required for each model to reproduce the observed
timescale.  The derived set of ejecta parameters are
($\Mej/\Msun$, $\KE/10^{51}$ erg) $=$ (2.7, 1.1), (4.4, 3.7),
(6.2, 8.4), (7.7, 13.0) and (12.5, 26.5) for the case of HE4, HE6,
HE8, HE10 and HE16, respectively.  The ejected
\Nifs\ mass is $\sim$0.07 $\Msun$ in all models.

Figure 3 shows the comparison between the observed and calculated
spectra. It seems that HE4, HE10 and HE16 are not consistent with SN
2008D, and that a model between HE6 and HE8 is preferable.

We thus conclude that the progenitor star of SN 2008D had a He core 
mass $M_{\alpha} =$ 6--8~$\Msun$ and exploded with $\Mej = 5.3 \pm 1.0$
$\Msun$ and $\KE = 6.0 \pm 2.5 \times 10^{51}$ erg.  The mass of the
central remnant is 1.6--1.8 $\Msun$, which is near the boundary
mass between the neutron star and the black hole.

\begin{figure}[t]
\begin{center}
\includegraphics[width=10.0cm]{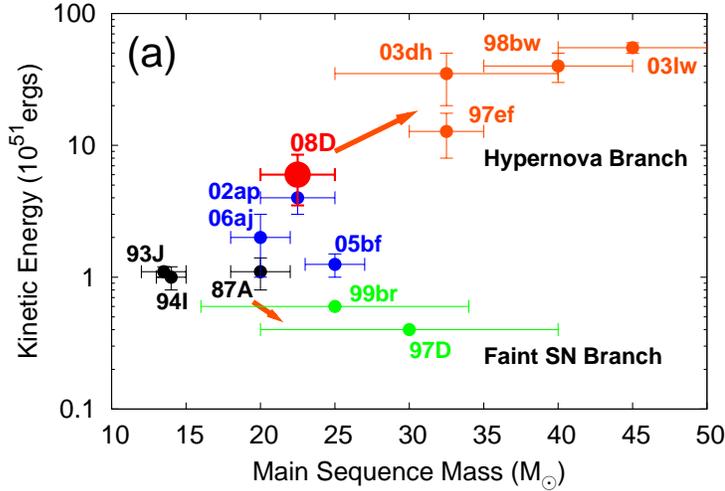}
\end{center}
\caption{
The kinetic explosion energy $\KE$ as a function of the main sequence
mass $\Mms$ of the progenitors for several supernovae/hypernovae
(\cite{tan09}).
}
\label{fig:ME}
\end{figure}

Figure \ref{fig:ME} shows $\KE$ as a function of $\Mms$
for several core-collapse SNe (see, e.g., \cite{nom07}).  SN
2008D is shown by a red circle.  Comparison with other SNe Ib is
possible only for SN 2005bf although SN 2005bf is a very
peculiar SN that shows a double peak LC
(\cite{anu05}; \cite{tom05}; \cite{fol06}; \cite{mae07}).
The LC of SN 2005bf is broader than that of SN 2008D,
while the expansion velocity of SN 2005bf is lower than that of SN
2008D.  These facts suggest that SN 2005bf is the explosion with lower
$\KE/ \Mej$ ratio.

The spectra of SN 2008D and Type Ib/c supernova (SN Ib/c) 1999ex
are very similar (\cite{val08}),
while SN 2005bf has lower He velocities.  The He lines
in Type IIb supernova (SN IIb) 1993J are very weak at this epoch.
The Fe features at 4500-5000\AA\ are similar in these four SNe,
but those in SN Ib 2005bf are narrower.
Malesani \etal\ (2008) suggested that the bolometric
LCs of SNe 1999ex and 2008D are similar.  The similarity in both the
LC and the spectra suggests that SN Ib/c 1999ex is also as energetic
as SN 2008D in the $\KE$--$\Mms$ diagram.

Malesani \etal\ (2009) also pointed the similarity of the LCs of SNe
IIb 1993J and Ib 2008D.  But the expansion velocity is higher in SN
2008D (see, e.g., \cite{pra95}).  Thus, both the mass and the kinetic
energy of the ejecta are expected to be smaller in SN IIb 1993J.
In fact, SN 1993J is explained by the explosion of a 4 $\Msun$ He
core with a small mass H-rich envelope (\cite{nom93}; \cite{sig94};
\cite{woo94}).

\begin{figure}[t]
\begin{center}
\includegraphics[width=10cm]{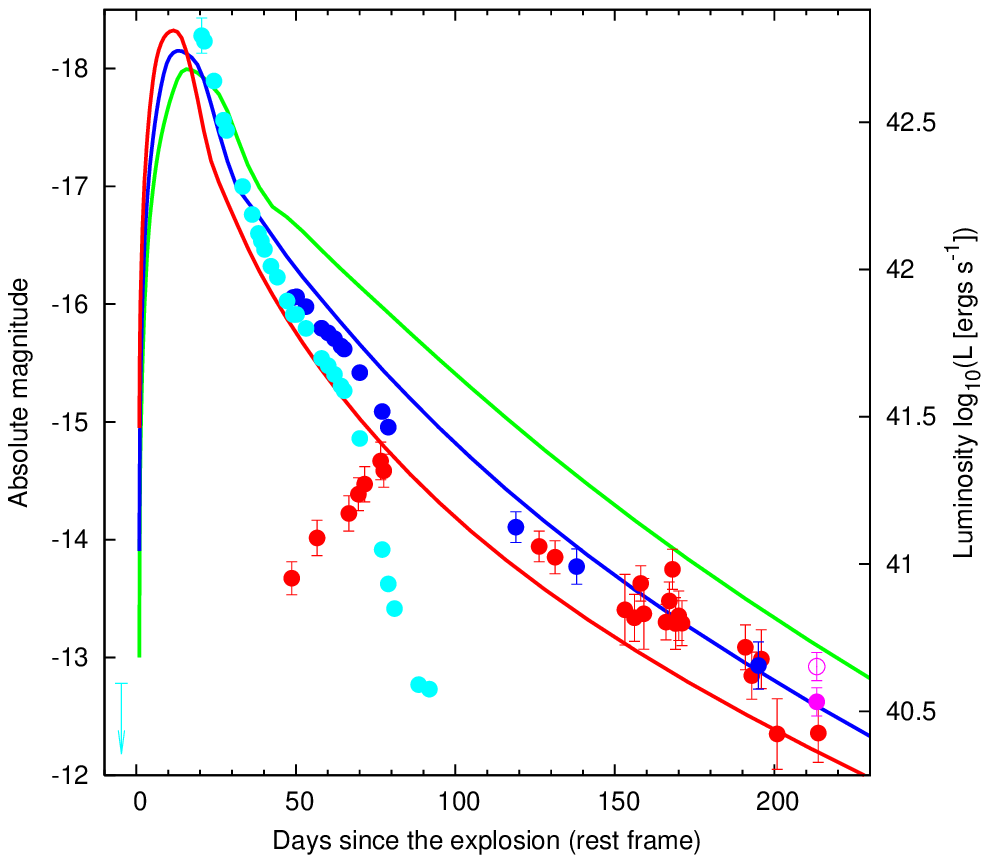}
\caption
{Comparison between the synthetic LCs for the models with $E_{51}=5$
and $\Mej=5.1$~$\Msun$, $E_{51}=10$ and $\Mej=4.9$ $\Msun$, and
$E_{51}=20$ and $\Mej=4.6$ $\Msun$, and the LCs of SN~2006jc
($L_{\rm UV}+L_{\rm opt}$, $L_{\rm IR,est}
(\nu<3\times10^{14}{\rm\ Hz})$, $L_{\rm bol}$, $L_{\rm IR,hot}
(\nu<3\times10^{14}{\rm\ Hz})$, $L_{\rm IR}
(\nu < 3\times10^{14}{\rm\ Hz})$) (\cite{tom08}).
}
\label{fig:LC}
\end{center}
\end{figure}

\section{Dust-Forming Type Ib Supernova SN 2006jc}

Another recent SN Ib 2006jc is characterized by the dust formation in
the ejecta as found from the near-infrared (NIR) and mid-infrared
(MIR) observations (e.g., \cite{smi08}; \cite{sak09}; \cite{mat08}).

We present a theoretical model for SN Ib 2006jc (\cite{tom08};
\cite{noz08}).  We calculate the evolution of the progenitor star,
hydrodynamics and nucleosynthesis of the SN explosion, and the SN
bolometric LC. The synthetic bolometric LC is compared with the
observed bolometric LC constructed by integrating the UV, optical,
NIR, and MIR fluxes.

The progenitor is assumed to be as massive as 40 $\Msun$ on the
zero-age main-sequence.  The star undergoes extensive mass loss to
reduce its mass down to as small as 6.9 $\Msun$, thus becoming a WCO
Wolf-Rayet star.  The WCO star model has a thick carbon-rich layer, in
which amorphous carbon grains can be formed. This could explain the
NIR brightening and the dust feature seen in the MIR spectrum.  We
suggest that the progenitor of SN~2006jc is a WCO Wolf-Rayet star
having undergone strong mass loss and such massive stars are the
important sites of dust formation.  We derive the parameters of the
explosion model in order to reproduce the bolometric LC of SN~2006jc
by the radioactive decays: the ejecta mass 4.9 $\Msun$, hypernova-like
explosion energy $10^{52}$ ergs, and ejected \Nifs\ mass 0.22 $\Msun$.

We also calculate the circumstellar interaction and find that a CSM
with a flat density structure is required to reproduce the X-ray LC of
SN~2006jc. This suggests a drastic change of the mass-loss rate and/or
the wind velocity that is consistent with the past luminous blue
variable (LBV)-like event.

We have thus found SN Ib 2006jc is almost a HN-like energetic
explosion.  This is suggestive for the SN Ib contribution to the
early enrichment in the Universe.  Also dust formation in WCO star 
seems to be quite important.

\section{Concluding Remarks}

We presented a theoretical model for SN 2008D associated with the
luminous X-ray transient 080109, which well reproduced the
bolometric LC and optical spectra.  This is the first detailed model
calculation for the SN Ib that is discovered shortly after the
explosion.  SN 2008D is located between the normal SNe and the
``hypernovae branch'' in the $\KE$--$\Mms$ diagram (upper panel of
Fig. \ref{fig:ME}).  The ejected \Nifs\ mass in SN 2008D ($\sim$0.07
$\Msun$) is similar to the \Nifs\ masses ejected by normal SNe and much
smaller than those in GRB-SNe.

These energetic SNe Ib, as indicated from both 2008D and 2006jc, and
also energetic SN Ib/c 1999ex, could result from the spiral-in of a
low mass binary companion into a massive star (\cite{nom95}).  The
spiral-in can eject the H-rich envelope by heating, and also bring the
orbital angular momentum into the core (\cite{nom01}).  SNe Ib are
less energetic than SNe Ic, because the effects of the spiral-in are
smaller in SNe Ib, where the He-layer is not ejected and smaller
angular momentum is brought into the core than in SNe Ic.  

Although the explosions are not as extreme as Hypernovae, such
energetic SNe Ib could make important contributions to the chemical
enrichment in early Universe.

\end{document}